\documentclass[11pt]{article}

\usepackage{amssymb,amsmath}
\usepackage[totalwidth=17cm,totalheight=24cm]{geometry}
\usepackage{graphicx}

\newcommand{\C}{{\mathbb C}}
\newcommand{\CP}{{\mathbb CP}}
\newcommand{\R}{{\mathbb R}}
\newcommand{\PT}{{\mathbb PT}}

\newcommand{\im}{{\rm i }}
\newcommand{\id}{{\mathbb I}}

\newcommand{\be}{\begin{eqnarray}}
\newcommand{\ee}{\end{eqnarray}}

\newtheorem{lemma}{Lemma}

\newcommand{\la}{\langle}
\newcommand{\ra}{\rangle}
\newcommand{\dir}{d\!\!\!\backslash}

\begin{document}

\title{Self-Dual Gravity}
\author{Kirill Krasnov \\ {\it School of Mathematical Sciences, University of Nottingham, NG7 2RD, UK}}

\date{v2: November 2016}
\maketitle
\begin{abstract}\noindent Self-dual gravity is a diffeomorphism invariant theory in four dimensions that describes two propagating polarisations of the graviton and has a negative mass dimension coupling constant. Nevertheless, this theory is not only renormalisable but quantum finite, as we explain. We also collect various facts about self-dual gravity that are scattered across the literature. 
\end{abstract}

\section{Introduction}

Self-dual gravity is a theory of gravity in four dimensions whose only solutions are Einstein metrics with a half of the Weyl curvature vanishing. Such metrics are also known in the literature as the gravitational instantons. Self-dual gravity is analogous to self-dual Yang-Mills theory, the latter being a four-dimensional gauge theory whose only solutions are connections with a half of the curvature vanishing. 

One purpose of this paper is to collect known facts about self-dual gravity. These are scattered across the literature spanning the last two decades. Some are reasonably well-known. Some other are contained in recent works of this author, in various degrees of explicitness, and are known less. We hope that collecting all these in one place will create a useful resource on the topic.

The other purpose of this text is to attract attention of the community to the fact that self-dual gravity is an interacting theory that describes two propagating degrees of freedom (two polarisations of the graviton), is diffeomorphism invariant, has a coupling constant of negative mass dimension, and in spite of all these trouble guaranteeing features gives rise to a perfectly well-behaved quantum theory. This theory is finite -- there are no quantum divergences. So, it gives the only known example of a consistent theory of quantum {\it pure} gravity in four dimensions. Superstring theory, when appropriately compactified, does give rise to a consistent theory of quantum gravity in four dimensions, but it has (infinitely) many extra fields. 

Of course one could object that self-dual gravity can only describe metrics of either Riemannian or split signatures (in Lorentzian signature vanishing of half of the Weyl implies vanishing of it all), and hence is not a physical theory.\footnote{Another not very physical feature of self-dual gravity is that many of its scattering amplitudes vanish, see below.} But one should recall that the argument of power counting non-renormalisability of GR has little to do with the metric signature, it is only based on the dimensionality of the coupling constant. In fact, all loop calculations that confirm non-renormalisability are done in the Euclidean setting. So, the striking fact here is that, as any other theory of gravity, self-dual gravity has a negative mass dimension coupling constant, but in spite of this manages to be not just renormalisable, but quantum finite. 

The mechanism for how this is possible is instructive and can be explained in simple terms already in the Introduction. First, we shall see that the Lagrangian of self-dual gravity treats two polarisations of the graviton on a different footing. Thus, let us choose our theory to describe metrics with vanishing self-dual (SD) part of Weyl. It would be more natural to call this theory anti-self-dual (ASD) gravity because it is the ASD part of Weyl that is allowed to be non-zero, but we shall continue, for brevity, to refer to it as SD theory. Then there is a polarisation of the graviton that respects the SD part of Weyl being zero condition. It is natural to refer to this polarisation as the negative one because this graviton will have only ASD part $W^-$ of Weyl non-zero. As we shall see below, the Lagrangian of SD gravity will be non-linear in the field that describes the negative helicity gravitons. It will however contain another field, used to describe the other (positive) helicity polarisation. The Lagrangian is linear in this field. It is then not hard to see that the theory described by a Lagrangian of this sort is one-loop exact. Thus, at tree level it is only possible to construct diagrams with one positive external leg and as many negative legs as one desires. At one-loop level it is only possible to have diagrams with negative external legs. It is not possible to construct diagrams with more than one loop.

The theory being one-loop exact, the study of quantum divergences, if any, reduces to that at one-loop. The one-loop effective action is given by the logarithm of the determinant of a certain first-order differential operator. This operator arises by linearising the theory around a fixed instanton background, and the effective action is a functional of the fields describing the background. For convenience of the computation this operator can be squared and then the determinant of the resulting second order operator found by the heat-kernel methods. As is usual in four dimensions, the logarithmic divergences, if any, are proportional to quantities constructed from the background curvature squared. One can then quickly convince oneself that on instanton backgrounds all possible quantities of this type reduce to topological invariants of the manifold. Thus, due to certain identities involving the curvature of instanton backgrounds, there are no possible counterterms and thus no divergences. This will be spelled out below.

The above argument can be compared to what happens with full GR at one loop \cite{'tHooft:1974bx}. In this case the possible logarithmic divergences are also integrals of curvature squared. One can then use the fact that the background is Einstein, as well as the identity expressing the Euler characteristic of the manifold as an integral of the curvature squared to eliminate all divergences apart from the divergence proportional to the total volume of the space. This is the reason why quantum gravity is one-loop finite in flat space (or renormalisable on a constant curvature background). The additional simplification that occurs in the self-dual case is that also the total volume of the manifold can be expressed as a certain linear combination of the Euler characteristics and the signature of the 4-manifold. So, to summarise, in SD gravity there are no divergences (of the type that can contribute to the S-matrix) at one loop, and there are no higher loops, so the theory is quantum finite. This should be contrasted with full GR, which is one-loop finite in flat space \cite{'tHooft:1974bx}, but requires the famous Goroff-Sagnotti counterterm \cite{Goroff:1985th} at two loops. 

The paper is organised as follows. One of the striking facts about the self-dual gravity is that it behaves in almost the same way as the more known self-dual YM theory. So, it is appropriate to start with a description of SDYM, which is what we do in Section \ref{sec:sdym}. We describe the covariant formulation of this theory, a convenient for computations gauge-fixing that makes the operator arising that of Dirac type, characterise the cubic interaction of this theory in terms of amplitudes, compute the Berends-Giele current and show that all tree-level amplitudes for more than 3 particles vanish on shell. We also discuss the vanishing of quantum divergences and the give the result for the one loop scattering amplitudes. We describe a relation to full YM, and sketch the twistor space description of SDYM. In Section \ref{sec:sdgr} we proceed with self-dual gravity. We present the material in precisely the same order, to highlight exact parallel between the two theories. 

We shall see that self-dual gravity behaves in exactly the same way as SDYM, the only difference being that SDGR is a bit more non-linear than SDYM (there is also a quartic vertex in gravity while the only interaction of SDYM is cubic), as well as the fact that SDGR has negative mass dimension coupling constant. Thus, self-dual gravity follows the same pattern that has been emerging for the relation between full gravity and YM: In spite of its non-renormalisable interaction, gravity is much more closely related to Yang-Mills theory than could be anticipated by inspecting the corresponding Lagrangians. On this point, we refer the reader to works of Bern and collaborators, e.g. \cite{Bern:2013yya}.

It is also worth pointing out that the formulation of SDGR on which this article is based and which exhibits the strongest analogy with SDYM is based on connections rather than metrics. The main rational for considering the connection rather than metric formulation is that the instanton condition, while second order in derivatives in the metric language (as a condition on the Riemann curvature), is a first order condition at the level of connections. The price to pay for this is that the connection formulation requires a non-zero cosmological constant (of arbitrary sign).

\section{Self-Dual Yang-Mills}
\label{sec:sdym}

We start this paper by describing what is known as the Chalmers-Siegel formulation \cite{Chalmers:1996rq} of self-dual Yang-Mills. We do this to showcase a strong analogy between self-dual Yang-Mills and self-dual gravity. 

SDYM in the covariant formulation \cite{Chalmers:1996rq} can be studied by usual Feynman diagram techniques, and we describe a very useful way of gauge-fixing this theory. The kinetic operator of the gauge-fixed theory is just an appropriate Dirac operator. This gauge-fixing was used in \cite{Krasnov:2015kva}, and also appears in the book \cite{Siegel:1999ew}, exercise VIB4.1c, in the context of the first order formalism for the full YM.  We sketch the computation of the Berends-Giele current \cite{Berends:1987me} and show that all tree level amplitudes (on a trivial background) with more than 3 external legs vanish on shell. We explain why there are no quantum divergences in this theory, and state the result for the one-loop scattering amplitudes. The non-vanishing such amplitudes are those involving only negative helicity gluons. 

Bardeen \cite{Bardeen:1995gk} has suggested that the close relationship between the self-dual and full YM, as well as the integrability of the former, may be of help to understand the latter theory. So far this idea has not been realised, possibly because no one has tried hard enough.

\subsection{The theory}

There are several non-covariant formulations of self-dual Yang-Mills, see \cite{Chalmers:1996rq} for a discussion of this. We will only present the covariant formulation, also from \cite{Chalmers:1996rq}. The Lagrangian of SDYM involves two fields. One is the usual YM connection field, the other can be referred to as the Lagrange multiplier field imposing the self-duality condition. The action is
\be\label{sdym}
S^{\rm SDYM}[A,B^+] = \int {\rm Tr} \left( B^+ \wedge F\right).
\ee
Here $F= dA + A\wedge A$ is the curvature 2-form, and $B^+$ is a Lie algebra valued 2-form field that is required to be self-dual
\be\label{B+}
B^+_{\mu\nu} = \frac{1}{2} \epsilon_{\mu\nu}{}^{\rho\sigma} B^+_{\rho\sigma}.
\ee
We work in either Riemannian or split signature. For both of these there is no imaginary unit in this formula. Trace everywhere stands for the matrix trace, with Lie algebra valued objects represented by matrices. The metric information enters (\ref{sdym}) via the requirement (\ref{B+}) that $B^+$ is self-dual. 

The field equations are as follows. Varying the action with respect to $B^+$ we get
\be\label{ym-feq}
F^+ =0,
\ee
which is correct field equation for SDYM. It says that the field strength is anti-self-dual, which then implies the usual YM field equation $d_A^\mu F_{\mu\nu}=0$. Connections satisfying (\ref{ym-feq}) are called instantons. Field configurations satisfying (\ref{ym-feq}) describe one (in our conventions negative) polarisation of the gluon. Varying with respect to the connection one gets
\be\label{ym-B-eqn}
d_A B^+=0.
\ee
Here $d_A$ is the covariant derivative with respect to the gauge field. We will later identify this as the field equation for the other polarisation of the gluon. 

\subsection{Linearisation}

The action (\ref{sdym}) can be expanded around an arbitrary instanton background. Thus, let the background connection satisfy (\ref{ym-feq}), while background $B^+=0$. Denote the perturbations by $b\equiv \delta B^+, a\equiv \delta A$. We then have the following linearised Lagrangian
\be\label{L2-ym}
{\cal L}^{(2)} = {\rm Tr}( b d_A a).
\ee
Here $d_A$ is the covariant derivative with respect to the background connection, and we have omitted the wedge symbol product for brevity. There is only the cubic interaction, and the interaction part of the Lagrangian reads
\be\label{L3-ym}
{\cal L}^{(3)} = 2 {\rm Tr}( b aa).
\ee

\subsection{Linearisation around a non-trivial $B^+$ background}

One can also take a more non-trivial background involving also a non-zero $B^+$ satisfying (\ref{ym-B-eqn}) with $A$ satisfying (\ref{ym-feq}). The interaction (\ref{L3-ym}) is unchanged, but the kinetic term changes to
\be\label{ym-L2-B}
{\cal L}^{(2)} = {\rm Tr}( b d_A a) + {\rm Tr}( B^+ aa),
\ee
which is a sort of a mass term for the connection. This extra term changes the propagator of the theory, see below.

\subsection{Spinor description}

We will only give details for the $B^+=0$ backgrounds, making some comments about more general case in \ref{ss-2.9}.

Given that $b$ is a self-dual two-form (with values in the Lie algebra), it is very convenient to write this object as a spinor. When this is done, one no longer has to keep in mind its self-dual property. Our spinor notations are explained in the Appendix.

We now translate (\ref{L2-ym}), (\ref{L3-ym}) into spinor notations. The connection perturbation field $a_\mu$ becomes an object $a_{MM'}$ (still Lie algebra valued). The self-dual 2-form field $b_{\mu\nu}$ becomes an object $b_{MN} \epsilon_{M'N'}$ with $b_{MN}=b_{(MN)}$. Thus, our self-dual field takes values in $S_+^2$. The free part of the Lagrangian becomes
\be\label{L2-ym-spin}
{\cal L}^{(2)} = {\rm Tr}\left(b^{MN} d_M{}^{M'} a_{NM'}\right).
\ee
Here $d_{MM'}$ is the operator of covariant derivative with respect to the background connection. The interaction becomes
\be\label{L3-ym-spin}
{\cal L}^{(3)} = 2 {\rm Tr}\left( b^{MN} a_{M}{}^{M'} a_{NM'}\right).
\ee
We have absorbed any numerical coefficient that arises in the spinor translation into $b^{MN}$.

\subsection{Gauge-fixing}

The main idea of the gauge-fixing procedure introduced below is to combine the field $b^{MN}$ with the auxiliary field in the ghost sector, to obtain a Dirac operator.  This was used in \cite{Krasnov:2015kva}, but also known to W. Siegel, see \cite{Siegel:1999ew}, exercise VIB4.1c.

The first-order differential operator in (\ref{L2-ym-spin}) maps $S_+\times S_- \to S_+^2$ and is thus degenerate. This degeneracy reflects the usual gauge invariance. To gauge fix this, we use the following gauge-fixing fermion
\be\label{ym-gauge}
\Psi = {\rm Tr}\left( \bar{c} \, d^{MM'} a_{MM'}\right) \equiv {\rm Tr}\left( \bar{c} \, \epsilon^{NM} d_M{}^{M'} a_{NM'}\right)
\ee
where $\bar{c}$ is the anti-ghost and as before $d_{MM'}$ is the background covariant derivative. This is the usual for YM theory gauge-fixing fermion, except that we are using the Landau gauge rather than the Feynman gauge. The gauge-fixing Lagrangian is then
\be\label{L-ym-gf}
{\cal L}_{gf} = {\rm Tr}\left( h_c \,\epsilon^{NM} d_M{}^{M'} a_{NM'}\right) + {\rm Tr}\left( \bar{c} \,d^{MM'} (d_{MM'} c + a_{MM'} c + c a_{MM'}) \right),
\ee
where $h_c$ is the ghost sector auxiliary field. As usual the full covariant derivative (including the perturbation field) has appeared in the ghost term. 

What is convenient about the gauge-fixing (\ref{ym-gauge}) is that the first term in (\ref{L-ym-gf}) can be combined with (\ref{L2-ym-spin}) by simply defining a new field 
\be
\tilde{b}^{NM} := b^{NM} + h_c \epsilon^{NM}.
\ee
This new field takes values in $S_+\times S_+$. Then (\ref{L2-ym-spin}) together with the first term in (\ref{L-ym-gf}) take the same form as (\ref{L2-ym-spin}) but with $\tilde{b}$ in place of $b$. We also remark that in (\ref{L3-ym-spin}) we can replace $b$ by $\tilde{b}$ if we add the symmetrisation on the $a$'s. We now drop the tilde on the $b$ for brevity, and write the full gauge-fixed Lagrangian as
\be\label{L-ym-full}
{\cal L}^{(2)}_{full} = {\rm Tr}\left(b^{NM} d_M{}^{M'} a_{NM'}\right)+ {\rm Tr}\left( \bar{c} \,d^{MM'} d_{MM'} c \right) , \\ \nonumber
{\cal L}^{(3)}_{full} =  2 {\rm Tr}\left( b^{MN} a_{(M}{}^{M'} a_{N)M'}\right) + {\rm Tr}\left( \bar{c} \,d^{MM'} (a_{MM'} c + c a_{MM'}) \right).
\ee
The operator that appears in the bosonic kinetic part is just the (covariant) Dirac operator
\be\label{dirac-ym}
\dir : S_+\times S_-\times {\mathfrak g}  \to S_+\times S_+ \times {\mathfrak g}.
\ee
Here ${\mathfrak g}$ is the Lie algebra. This operator is elliptic and all the gauge has been fixed. 

\subsection{Amplitudes}

A useful exercise is to characterise the Lagrangian of SDYM in terms of amplitudes. For doing this we pass to the split $(++--)$ signature where null momenta as well as self-dual field configurations are possible. In this signature all spinor objects are real. We will also work around the trivial background connection (as well as $B^+=0$), as is appropriate in the amplitude context. 

The linearised (gauge-fixed) field equations are
\be\label{lin-feq-ym}
\partial_{M}{}^{M'} a_{NM'} = 0, \qquad \partial^M{}_{N'} b_{NM} =0.
\ee
Note that these are just massless Dirac equations for "spinors" valued in $S_+\times S_-$ and $S_+\times S_+$ respectively. The Lie algebra index decouples in this linearised case. The Dirac operator acts on the second spinor index in each case. 

Applying the Dirac operator one more time and using the fact that the Dirac squared is the Laplacian we see (passing to the momentum space) that field equations imply that the momentum vector must be null. Such vectors, when written in spinor form, are products of two spinors of different types
\be\label{k2}
k^2=0 \Leftrightarrow k_{MM'} = k_M k_{M'}.
\ee
Here $k_M, k_{M'}$ are real and independent (in the case of split signature) spinors. We can then immediately construct the polarisation vectors solving (\ref{lin-feq-ym}). The negative helicity polarisation vector is given by
\be\label{eps-}
\epsilon^-_{MM'}(k) = \frac{q_M k_{M'}}{\la q k\ra}.
\ee
Here $q_M$ is an arbitrary reference spinor, whose presence here reflects the possibility of making gauge transformations. Physical amplitudes are $q$-independent. The denominator is necessary to make this polarisation spinor homogeneous degree zero in $q$. The general solution of the first equation in (\ref{lin-feq-ym}) is a linear combination of plane waves weighed with such polarisation tensors. It is most convenient to take the reference spinor $q$ to be the same for all negative helicity gravitons. The polarisation spinor in (\ref{eps-}) is dimensionless. 

Similarly, for the positive polarisation, the general solution of the second equation in (\ref{lin-feq-ym}) is a combination of plane waves with the polarisation spinor
\be
\epsilon^+_{MN}(k) = k_M k_N.
\ee
This polarisation spinor has mass dimension one, as is appropriate for a field of mass dimension two.

We now evaluate the first term in the second line of (\ref{L-ym-full}) on shell. Thus, we take the states $1,2$ to be of negative helicity, and $3$ to have positive helicity. We get for the colour-ordered 3-point amplitude
\be\label{A3-ym-1}
A^{--+} = \frac{ \la 3q\ra^2 [12]}{\la 1q\ra \la 2q\ra},
\ee 
where our notation is that $(k_1)_{AA'} = (k_1)_A (k_1)_{A'} \equiv 1_A 1_{A'}$, the angle bracket stands for the contraction of unprimed and the square bracket of primed spinors. Using the momentum conservation $11'+22'+33'=0$ to eliminate the reference spinor $q_A$ we get the familiar formula
\be\label{A3-ym-2}
A^{--+} = \frac{[12]^3}{[13][32]}.
\ee
This is the only 3-point amplitude in this theory. As we shall see below, this is also the only non-zero amplitude at tree level (on the trivial background).

\subsection{Berends-Giele current}

We now compute the colour-ordered Berends-Giele current, which is defined as the sum of all colour-ordered Feynman diagrams with all but one leg on-shell. A convenient rule is that the off-shell leg is taken {\it with} the propagator on that leg. We take the on-shell legs to be those of negative polarisation, as this gives the most interesting current. 

The one-point current is just the polarisation tensor (\ref{eps-}) itself. Anticipating what will happen, we write the general current as
\be\label{current-ym}
J_{MM'}(1,\ldots,n) = q_M q^E (\sum_{i=1}^n k_i)_{EM'} J(1,\ldots,n),
\ee
where $J(1,\ldots,n)$ is now a scalar. We have 
\be
J(1)=\frac{1}{\la q1\ra \la 1q\ra}.
\ee

The second current is computed putting two order one currents (polarisations) into the cubic vertex and applying the final leg propagator. It is easy to see that this gives
\be
J(1,2) = \frac{1}{\la q1\ra \la 12\ra \la 2q \ra},
\ee
where we wrote the result in a suggestive form.

To compute the third current we note that in forming a colour-ordered amplitude we can either attach the gluon number 3 to the current $J_{MM'}(1,2)$ or can attach gluon number 1 to the current $J_{MM'}(2,3)$. This gives
\be\nonumber
J(1,2,3) = \frac{1}{\la 1q\ra\la 2q\ra\la 3q\ra}\left( \frac{ \la q1\ra [13] + \la q2\ra [23]}{\la 12\ra} + \frac{ \la q2\ra [12] + \la q3\ra [13]}{\la 23\ra} \right)\frac{1}{\la 12\ra [12] +\la13\ra [13]+\la 23\ra [23]}.
\ee
We can then use Schouten identity in the form
\be
\la q1\ra \la 23\ra +\la q3\ra \la 12\ra = \la q2\ra\la 13\ra
\ee
to notice that the numerator cancels the propagator in the denominator and we have
\be
J(1,2,3)= \frac{1}{\la q1\ra\la 12 \ra\la 23\ra \la 3q\ra}.
\ee
The pattern is becoming clear. It is not hard to write a recursion relation \cite{Berends:1987me} for the general current and prove that
\be\label{Jn}
J(1,\ldots,n) = \frac{1}{\la q1\ra \la 12\ra \la 23\ra \ldots \la (n-1) n\ra \la nq\ra}.
\ee
This is essentially the famous Parke-Taylor formula \cite{Parke:1986gb} for the so-called MHV amplitudes. We thus see that the basic ingredient of this formula arises already in self-dual YM theory. This was noted by several authors, notably by \cite{Bardeen:1995gk}. 

The form of the current (\ref{current-ym}) together with the result (\ref{Jn}) immediately shows that all the on-shell amplitudes with more than 3 gluons are zero. Indeed, first of all, we already know that at tree-level there can only be amplitudes with at most one positive leg. So, let us compute these amplitudes by first computing the current with all negative on-shell legs, and then putting the off-shell leg on-shell inserting in it a positive polarisation gluon. In the process of doing this we must multiply the current by the propagator that by our convention was always included in the off-shell leg. However, since the current does not have any pole to cancel this propagator, and the propagator is zero on-shell, the whole on-shell amplitude is rendered zero. Thus, on the trivial background $B^+=0,A=0$ all tree-level amplitudes with $n>3$ are zero.

In order to avoid confusion we emphasise that, while the Berends-Giele current in SDYM is non-trivial and is given by (\ref{Jn}), the tree-level amplitudes in SDYM are trivial (zero). In particular, in spite of the fact that one sees in (\ref{Jn}) all the ingredients of the MHV amplitude Parke-Taylor formula, there are no non-zero $(++-\ldots -)$ MHV amplitudes in SDYM. The reason why the current (\ref{Jn}) resembles the Parke-Taylor formula so close is that one can take the limit of the Parke-Taylor formula in which the momenta of two of the positive helicity gluons become colinear. This limit must be proportional to the current (\ref{Jn}), and this is why the two formulas are so closely related. 

At first sight, it may seem that more amplitudes can be non-zero on a more general $B^+\not=0$ background. Thus, one can continue to take $A=0$, but take a non-zero $B^+$ solving $\partial_{A'}{}^B B^+_{BA}=0$. In this case, there is a second, mass-like term in the linearised Lagrangian (\ref{ym-L2-B}). This changes the propagator in that there is now also the $\la bb\ra$ part of the propagator. There are now more diagrams that can be constructed at tree level. In particular, it may seem that the all negative helicity tree level scattering amplitudes do not have to vanish anymore. However, this is an illusion, because we can equally well treat this correction to the propagator as a new $B^+ aa$ vertex. This vertex can connect two Berends-Giele currents computed by the procedure described above. However, when we attempt to put $B^+$ on shell the result will vanish, as we already know from the previous discussion. So, there are no new tree-level amplitudes one can generate on  $B^+\not=0$ backgrounds, provided of course that $B^+$ satisfies its field equation. 

\subsection{Quantum theory}

The fact that one of the fields, namely $B^+$, enters the Lagrangian linearly immediately tells us that the theory is one-loop exact. This is because at tree level one can only construct diagrams with at most one external $B^+$ leg. The propagator of the theory takes $B^+$ into $A$, and so at one loop only diagrams with no external $B^+$ leg can be formed. No higher loop diagrams exist. 

To study the theory at one loop one can use the background field method formalism, and evaluate the path integral of the theory linearised around an arbitrary background. We consider in details the case of a $B^+=0$ background, but the general background can be treated by the same method. We make some remarks on this at the end. We follow the computation in \cite{Krasnov:2015kva}, which in our case of self-dual theory becomes much simpler. 

The action to study is (\ref{L2-ym}). Once gauge-fixed the problem reduces to computing the determinant of the operators appearing in the first line in (\ref{L-ym-full}). We perform this computation by passing to a second-order problem. Thus, we first rewrite the bosonic part of the linearised Lagrangian as
\be\label{dirac-form}
{\cal L}^{(2)} = \frac{1}{2} \left( b \quad a  \right) \left( \begin{array}{cc} 0 & \dir \\ \dir^* & 0 \end{array} \right) \left( \begin{array}{c} b \\ a\end{array}\right)\equiv \frac{1}{2} \left( b \quad a  \right) D\left( \begin{array}{c} b \\ a\end{array}\right)
\ee
Here 
\be
\dir: V\times S_- \to V\times S_+,  \qquad 
(\dir a)_{NM} := d_N{}^{M'} a_{MM'}, 
\ee
is the chiral part of the Dirac operator and 
\be
\dir^*: V\times S_+ \to V\times S_-, \qquad 
(\dir^* b)_{MM'} := d_{M'}{}^N b_{MN}, 
\ee
is its adjoint operator. The operator $D$ in (\ref{dirac-form}) is then the usual Dirac operator acting on 4-component spinors, here with values in $V$
\be
V:= S_+\times {\mathfrak g},
\ee
The one-loop (and thus all loop) effective action is then given by
\be
S_{eff} = -\frac{1}{2} \log{\rm det}(D) + \log{\rm det}(d^2),
\ee
where the last term is the contribution from ghosts, i.e. the second term in the first line in (\ref{L-ym-full}).
To compute the first determinant we square the Dirac operator
\be\label{s-eff}
S_{eff} = -\frac{1}{4} \log{\rm det}(D^2) + \log{\rm det}(d^2),
\ee
with
\be\label{dirac-square}
D^2=\left( \begin{array}{cc} \dir^* \dir & 0 \\ 0& \dir \dir^* \end{array} \right)
\ee
being an operator of Laplace type. 

\subsection{Absence of quantum divergences}
\label{ss-2.9}

Logarithmic quantum divergences, if any, can now be computed by the standard heat kernel techniques, see e.g. \cite{Vassilevich:2003xt}. For the operator $D^2$ this computation has been performed in \cite{Krasnov:2015kva}, in a much more involved set-up. But the final result of relevance for us here could be written without any calculation. Indeed, the final result must be an integral of the background curvature squared, because only such terms arise in the heat kernel expansion at this order. We can then use the fact that 
\be\label{inst-number}
\int {\rm Tr}(F\wedge F) = \frac{1}{4} \int {\rm Tr}\left( (F_{MN})^2 - (F_{M'N'})^2\right)
\ee
is a total divergence that does not contribute to S-matrix. This quantity only depends on the fibre bundle, and (an appropriate multiple of it) is known as the first Pontryagin number. This means that modulo surface terms we can replace integrals of the ASD part of the curvature squared $(F_{M'N'})^2$ with integrals of the SD part of the curvature squared $(F_{MN})^2$. But on instanton backgrounds $F_{MN}=0$ by definition, and so the integral of the full curvature squared reduces on instanton backgrounds to the topological term (\ref{inst-number}) that has the interpretation as measuring the number of instantons. The one-loop divergence is thus proportional to the instanton number, and does not contribute to the S-matrix. The theory is quantum finite. 

The coefficient in front of the divergence proportional to the instanton number can be extracted from results in \cite{Krasnov:2015kva}, see in particular the formula (66) of this reference, with $A=B=0$. We refrain from giving these unnecessary details here. 

Let us now discuss what seems to be much more non-trivial case of a background with $B^+\not=0$. In this case there is a second term in the linearised Lagrangian, see (\ref{ym-L2-B}). The computation is still possible following the same idea, namely squaring the arising first-order operator and then using the heat kernel technology. This computation is possible, and in fact has been done (even in greater generality) in \cite{Krasnov:2015kva}. The result of interest for us here can be extracted from the formula (66) of \cite{Krasnov:2015kva} by putting the matrix $A$ to zero. Note that in the notations of that paper this matrix encodes terms quadratic in $b$. In particular, it has nothing to do with the connection. The result in \cite{Krasnov:2015kva} shows that, while terms of the $(B^+)^2$ type are present, they always appear together with factors of $A$. So, setting the matrix $A$ to zero also eliminates all $B^+$ dependent terms. 

The fact that the one-loop effective action on a $B^+\not=0$ background does not depend on $B^+$ can also be understood without any computation.\footnote{The author is grateful to Dmitri Bykov for a discussion that clarified this point.} Indeed, if there was some non-trivial dependence, we could expand the result in power series in $B^+$. We would then obtain one-loop amplitudes with one or more external $b$ legs. As we know from the preceding discussion, this is not possible. So, we learn from this argument that one-loop effective action cannot depend on $B^+$. It can therefore be computed on $B^+=0$ background, which is the much simpler computation reviewed above. The observed in (66) of \cite{Krasnov:2015kva} fact that for $A=0$ the dependence on $B^+$ drops out is thus not a miracle, it must have happened if the calculation was done correctly. 

To conclude, even though a computation with $B^+\not=0$ is possible via the same technology, this much more involved computation is not needed, as the result cannot depend on $B^+$. The theory is proved finite by a simple argument on a $B^+=0$ background. In particular, no computation is required, possible divergences cannot contribute to the S-matrix by a simple argument involving the Pontryagin number. 

\subsection{One loop amplitudes}

As we have just seen, there are no one loop divergences in SDYM (and no higher loops). However, the finite parts of one loop integrals are non-zero. Historically, people first computed one loop amplitudes in full YM. It was then observed that for special helicity configurations, namely all minus in our conventions, the full YM amplitudes coincide with those obtained directly within SDYM. 

The relevant references are as follows. The form of the same helicity one loop $n$-gluon amplitude was conjectured in \cite{Bern:1993qk} (see also reference [5] of that paper) using the collinear limit arguments. Around the same time Mahlon \cite{Mahlon:1993fe} used the technology \cite{Berends:1987me} of recursion relations to compute one loop multi-photon amplitudes in QED. Mahlon then applied the same recursive technology to compute a multi-gluon amplitude from a quark loop \cite{Mahlon:1993si}. Using supersymmetry, one can show that the same results apply to pure QCD amplitudes. In particular, the one loop all minus 4-point amplitude is a multiple of
\be\label{all-minus-ym}
A_{1-loop}^{----} \sim \frac{[12][34]}{\la 12\ra \la34\ra}.
\ee
Note that this is a purely rational result that does not have any cuts. This is related to the fact that the on-shell amplitudes that could appear on the cut according to the Cutkosky rules are absent. A closed form expression for an arbitrary number of negative helicity particles is also known, and is given in \cite{Bern:1993qk}.

Cangemi \cite{Cangemi:1996rx}, \cite{Cangemi:1996pf} and then Chalmers and Siegel \cite{Chalmers:1996rq} then showed that the same helicity one loop amplitudes in SDYM are the same as those in full YM. Cangemi's argument was an explicit computation, but using the action that differs from (\ref{sdym}). The argument in \cite{Chalmers:1996rq} is more elementary, and follows from the relation between the full and SD YM Feynman rules, see below.

Bardeen \cite{Bardeen:1995gk} suggested that the non-vanishing of the one loop all-minus amplitudes in the SDYM theory may be related to an anomaly in the conservation of the symmetry currents responsible for the integrability of the theory. So far, to the best of our knowledge, this suggestion has not been verified. 

It would be interesting to compute (\ref{all-minus-ym}) directly from (\ref{sdym}) using the Feynman rules described above. This can also shed light on the suggested \cite{Bardeen:1995gk} anomaly interpretation of the result (\ref{all-minus-ym}).

\subsection{Relation to the full Yang-Mills}

The relation to the full YM theory arises by writing the YM action in the following form
\be\label{YM}
S^{YM} = \int {\rm Tr}\left( B^+ \wedge F - g^2 B^+\wedge B^+\right).
\ee
Integrating out the auxiliary field $B^+$, which, unlike in the SDYM case, is possible, one gets the YM action in the form of the integral of the $(F^+)^2$, which agrees with $F^2$ modulo a surface term. 

The quantum theory for (\ref{YM}) is constructed completely analogously to what we have done for the self-dual case. The gauge-fixing is performed in an analogous way, except that now the Feynman gauge is more convenient, see \cite{Krasnov:2015kva} where it is explained that in this case it is also possible to absorb the auxiliary field for the ghosts into the $B^+$ field perturbation. The interaction is also unchanged as compared to the SDYM case. The only difference is that now the kinetic term has an addition contribution $g^2 b^2$ in it. This changes the propagator, and apart from the propagator connecting the $b$ and $a$, there now is also a propagator connecting $a$ with $a$. So it becomes possible to construct many more diagrams. In particular, the theory is no longer one loop exact. 

At the level of the amplitudes, the field $a$ can now describe gluons of both helicities, but $b$ continues to describe only the positive helicity. So, when computing the full YM amplitudes involving $n$ negative helicity gluons, all polarisation spinors must be inserted into the $a$ legs of the vertices. At the same time, the structure of the Feyman rules is such that at tree level there is at least one $b$ external leg. At one loop level, the minimal number of external $b$ legs is zero. These tree level and one loop diagrams with the minimal possible number of external $b$ legs are the same as in SDYM theory. It is then clear that the one loop all negative amplitudes in full YM are those that do not have any $b$ external legs, and are therefore the same as in the SDYM. 

\subsection{Twistor space description}

The integrability of SDYM theory is most easily seen in its twistor description, in which the non-linear SDYM equations are expressed as the compatibility condition of a certain linear system, see e.g. \cite{Ablowitz:1993ec}.

A twistor action for SDYM has been described in \cite{Witten:2003nn} and then in \cite{Mason:2005zm}. We review only the very basics of this description. The twistor space over the conformal compactification $S^4$ of $R^4$ is the 2-component spinor bundle over $S^4$. The projective twistor space is then ${\mathbb PT}=\CP^3$. The connection allows to define a $(0,1)$ Lie algebra valued form $a$ on $\CP^3$. The field $b$ gets represented as a $(0,1)$-form that is Lie algebra and ${\cal O}(-4)$ valued. The lift of (\ref{sdym}) to the twistor space is
\be\label{sdym-twistor}
S[a,b] = \int_{\mathbb PT} {\rm Tr}(b\wedge f) \wedge \Omega.
\ee
Here $f=\bar{\partial} a +a\wedge a$ is the $(0,2)$ part of the curvature of the connection, and $\Omega$ is the $(3,0)$ holomorphic top form on $\CP^3$. It is of homogeneity degree ${\cal O}(4)$, which makes the integrand in (\ref{sdym-twistor}) homogeneity degree zero and the integral well-defined. Integrating the above action over the fiber we get (\ref{sdym}) with the self-dual 2-form $B^+$ arising as the Penrose transform of $b$
\be
B^+ = b_{AB} \Sigma^{AB}, \qquad 
b_{AB} =\int_{\CP^1} \pi_A \pi_B b\wedge \pi^C d\pi_C.
\ee
We refer the reader to \cite{Mason:2005zm} for more details. 

\subsection{Colour/kinematics duality}

The purpose of this subsection is to mention that the colour/kinematics duality \cite{Bern:2008qj} that is known to be true in full YM theory can be shown to hold in self-dual YM by analysing Feynman rules. This was first observed in \cite{Monteiro:2011pc} in a non-covariant version of the theory. A covariant version of the argument is also possible. The computation that leads to this conclusion is very similar to the computation of the Berends-Giele current, except that in this case it is more natural to consider the full (not colour-ordered) amplitudes. Details of this are given in  \cite{Fu:2016plh} and will not be repeated here. It would be interesting to try to understand the statement gravity equals YM squared that is an application of the colour/kinematics duality in the self-dual setup and from the point of view of Feynman diagrams. We leave this for future research.

\section{Self-Dual Gravity}
\label{sec:sdgr}

We are now ready to proceed with similar constructions in the case of self-dual gravity (SDGR). What we mean by self-dual gravity is a theory of gravity in four dimensions whose equations force metrics to be (i) Einstein $R_{\mu\nu}=\Lambda g_{\mu\nu}$; (ii) have one of the two chiral halves of the Weyl curvature vanishing, say $W^+_{\mu\nu\rho\sigma}=0$. 

There are several non-covariant formulations of SDGR available in the literature. In particular, Plebanski \cite{Plebanski:1975wn} gave a description of gravitational instantons in the zero scalar curvature case. Plebanski equation is nicely reviewed in \cite{Monteiro:2011pc}. This work also contains references to papers describing the lightcone action for self-dual gravity.

There is a covariant description of SDGR in \cite{Siegel:1999ew}, exercise IXA5.6, see also \cite{Siegel:1992wd} for a supersymmetric version. This description can be obtained by taking the "chiral" first order action for GR, and rescaling the fields so as to remove the $AA$ term from $F=dA+AA$. This modifies the local gauge invariance of the chiral GR action from the chiral half ${\rm SO}(3)$ of Lorentz group to ${\rm U}(1)^3$. This author doesn't know if a covariant gauge-fixing of this action is possible.  

In this paper we describe a different covariant formulation. It was known for a long time that gravitational instantons (of non-zero scalar curvature) can be described very economically via ${\rm SO}(3)$ connections rather than metrics. The first reference on this we are aware of is the paper by Gindikin \cite{Gindikin},\cite{Gindikin:1982kk}. These papers used ${\rm SL}(2)$ connections rather than ${\rm SO}(3)$, but contain the key idea of all later descriptions. The connection description was rediscovered by Capovilla, Dell and Jacobson in \cite{Capovilla:1990qi}, in the context of their work \cite{Capovilla:1989ac} on "General Relativity without the metric". It was again rediscovered in \cite{Fine}. Another relevant set of papers is that by Torre \cite{Torre:1990zb}, \cite{Torre:1990ux}, \cite{Torre:1991sj}. Here the author describes the linearisation of the instanton condition in the connection language, as also we do below. The connection description of instantons also appears in \cite{Capovilla:1992ep}. 

Most of the above authors consider equations describing gravitational instantons in the language of connections rather than an action that gives these equations as critical points. So, it is not easy to pinpoint the first reference that contains this action. Given that, once the equations are known, the action with the right properties presents itself almost immediately, we restrict ourselves to just one recent reference \cite{Herfray:2015fpa} explicitly containing the action. 

One of the main attractive features of the description we are to present is that the instanton condition is a first order PDE on the basic field. This is similar to the YM case, but should be contrasted with the metric description in which the instanton condition, being that on the curvature, is a second order PDE. The other attractive feature is that in this description the gauge-fixing is based on a nice "instanton deformation complex", see the discussion around (\ref{def-complex}) below. 

\subsection{The action}

Similar to (\ref{sdym}) there are two types of fields in the action. One is the Lagrange multiplier field in which the Lagrangian is linear. It will describe one of the two polarisations of the graviton. The field in terms of which the action is non-linear will be an ${\rm SO}(3)$ connection field. We will explain its relation to the metric below. The action reads
\be\label{sdgr}
S^{\rm SDGR}[A,\Psi] = \int \Psi^{ij} F^i \wedge F^j.
\ee
Here $F^i = dA^i + (1/2) \epsilon^{ijk} A^j\wedge A^k$ is the curvature 2-form, and $\Psi^{ij}=\Psi^{(ij)}$ is the Lagrange multiplier field that is required to be tracefree $\Psi^{ij}\delta_{ij}=0$. The action (\ref{sdgr}) is clearly ${\rm SO}(3)$ and diffeomorphism invariant. Note that no metric appears or is used in the construction of the action.

The Euler-Lagrange equations following from (\ref{sdgr}) are as follows. By varying with respect to the Lagrange multiplier field we get
\be\label{feqs-ff}
F^i \wedge F^j \sim \delta^{ij}.
\ee
This equation says that the 4-form valued matrix $F^i\wedge F^j$ has vanishing tracefree part, and thus has only the trace part proportional to $\delta^{ij}$. The proportionality coefficient follows by computing the trace of both parts of this equation. It is this equation that was observed in \cite{Gindikin}, \cite{Capovilla:1990qi} to describe the gravitational instantons, as we shall review below. Note that (\ref{feqs-ff}) is a first-order differential equation on the basic (connection) field.  

The equation one obtains by varying with respect to $A$ is 
\be
d_A \left( \Psi^{ij} F^j\right) =0.
\ee
Below we will see that this will describe one of the two polarisations of the graviton propagating in this theory.

We also note that one could use a more general action than (\ref{sdgr}), allowing an arbitrary symmetric matrix in front of $F^i\wedge F^j$, but requiring the trace of this matrix to be a constant, so that the extremisation is only carried out with respect to the tracefree part. The part of the action proportional to the trace part then gives rise to the Pontryagin number and does not change the field equations. This observation is related to the formalism for the connection description of gravity described in \cite{Herfray:2015fpa}.

\subsection{2-forms in four dimensions}

We now need to explain why (\ref{feqs-ff}) describes instantons. To explain this, we need to start by reviewing the geometric fact that a generic triple of 2-forms in four dimensions defines a metric. 

Thus, let $B^i$ be a triple of two-forms in four dimensions. First, we need to spell out the notion of genericity. A triple $B^i$ of 2-forms is called {\it generic or non-degenerate} if the $3\times 3$ matrix of wedge products $B^i\wedge B^j$ has non-zero determinant. Here, to check the condition one divides the 4-form valued matrix $B^i\wedge B^j$ by an arbitrarily chosen volume form. Clearly, the non-degeneracy condition does not depend on which volume form is used for this purpose. We assume that the manifold is orientable, so that a globally defined volume form exists. 

Another natural notion arising in this context is that of {\it definite} triples $B^i$. A non-degenerate triple of 2-forms $B^i$ is called definite if the matrix of wedge products $B^i\wedge B^j$ is definite, i.e. has eigenvalues of the same sign. Once again, to check definiteness one divides $B^i\wedge B^j$ by an arbitrarily chosen volume form $v$ and checks definiteness of the arising symmetric $3\times 3$ matrix $B^i\wedge B^j/v$. 

A definite triple $B^i$ defines a natural {\it orientation}. This is the orientation given by the volume form $v$ such that the matrix $B^i\wedge B^j/v$ is positive definite. 

Now, let $B^i$ be a definite triple of 2-forms. It defines a metric by the following formula 
\be\label{Urbantke}
(\xi,\eta)_g {\rm (vol)}_g = \frac{1}{6} \epsilon^{ijk} i_\xi B^i \wedge i_\eta B^j \wedge B^k.
\ee
Here on the left we have the metric contraction of two vector fields $\xi,\eta$ multiplied by the volume form for this metric. The volume form is chosen in the same orientation as that defined by $B^i$. On the right we have the top form constructed from the triple of 2-forms $B^i$. The volume form ${\rm (vol)}_g$ can be computed by taking the determinant of both sides. This shows that the volume form is homogeneity degree 2 in $B^i$, and the metric itself is homogeneity degree one in $B^i$. The ultimate origin of the formula (\ref{Urbantke}) is the geometry of 3-forms in 7 dimensions, see \cite{Krasnov:2016wvc}, but we will deviate from the main topic too far if we are to explain this.  

It can be shown that the metric (\ref{Urbantke}) is non-degenerate (has non-zero determinant) whenever $B^i$ is non-degenerate. Moreover, this metric is of signature all plus or all minus when the matrix of wedge products $B^i\wedge B^j$ is definite. 

The geometrical meaning of the formula (\ref{Urbantke}) is that this metric makes the definite triple $B^i$ {\it self-dual} with respect to the Hodge star operator defined by this metric, in the orientation defined by $B^i$. 

The formula (\ref{Urbantke}) also defines a sign of the triple $B^i$. Thus, when the metric (\ref{Urbantke}) is of signature all plus we call $B^i$ to be {\it positive definite}, and when we get the signature all minus we call $B^i$ to be {\it negative definite}. It is clear that multiplying the positive definite triple by the minus sign one gets a negative definite triple. The sign of the triple will be identified with (minus) the sign of the cosmological constant. 

For completeness we also mention that when $B^i$ is real non-degenerate but indefinite then (\ref{Urbantke}) gives a metric of split signature. It is also possible to get Lorentzian signature metrics out of (\ref{Urbantke}), but this requires complex $B^i$ satisfying reality conditions $B^i\wedge (B^j)^*=0$ together with the reality condition on the determinant of the matrix $B^i\wedge B^j$. 

All the statements in this subsection are those of linear algebra. They are proved by noticing that for a non-degenerate $B^i$ there is always a ${\rm GL}(3)$ matrix $G^{ij}$ such that $B^i = G^{ij} \Sigma^i$, where $\Sigma^i$ satisfy an orthonormality property. Thus, in the case of positive definite $B^i$ the matrix $G^{ij}$ can be taken to be the positive branch of the square root of the positive definite matrix $B^i\wedge B^j$, and then $\Sigma^i \wedge \Sigma^j \sim \delta^{ij}$. One then constructs the complex linear combinations $\Sigma^\pm = \Sigma^1 \pm \im \Sigma^2$ that are simple $\Sigma^\pm\wedge \Sigma^\pm=0$ and thus decomposable. Therefore there exist complex one-forms $u,v$ such that $\Sigma^+ = u\wedge v, \Sigma^-=\bar{u}\wedge \bar{v}$. The bar here denotes the complex conjugation. Because $\Sigma^+\wedge \Sigma^- \not=0$ the 4 one-forms $u,v,\bar{u},\bar{v}$ form a frame. The data in $\Sigma^\pm$ is not enough to fix this frame completely because one can apply an ${\rm SL}(2,\C)$ transformation to $u,v$ without changing $\Sigma^\pm$. But using the fact that $\Sigma^3\wedge \Sigma^\pm=0$ as well as $\Sigma^3\wedge \Sigma^3=(1/2)\Sigma^+\wedge \Sigma^-$ together with the ${\rm SL}(2,\C)$ freedom in choosing $u,v$ we can achieve $\Sigma^3=(\im/2)(u\wedge\bar{u}+ v\wedge\bar{v})$. This fixes the frame completely. The metric (\ref{Urbantke}) is then in the conformal class of $ds^2=u\otimes\bar{u}+v\otimes\bar{v}$. The conformal factor is $({\rm det}(G))^{1/3}$.  All statements above follow from this construction. 

The case of split and Lorentzian signatures is proved using the same basic idea - one writes $B^i$ as a ${\rm GL}(3)$ matrix acting on $\Sigma^i$ with desired properties, and then these $\Sigma^i$ define some decomposable forms which in turn define the metric with respect to which $\Sigma^i$ are self-dual. 

\subsection{The connection description of instantons}

We now apply (\ref{Urbantke}) to the curvature 2-forms. The claim is that metrics (\ref{Urbantke}) with $B^i=F^i$, where $F^i$ satisfy (\ref{feqs-ff}) are anti-self-dual Einstein with non-zero cosmological constant. The sign of the cosmological constant is negative of that of the triple $B^i$. 

This claim is proved in several steps. The first step is the following lemma
\begin{lemma} Let $B^i$ be a definite triple of 2-forms. Then the equation
\be\label{compat}
d_A B^i  \equiv dB^i + \epsilon^{ijk} A^j \wedge B^k =0
\ee
viewed as an algebraic equation for the connection $A^i$, has a unique solution. 
\end{lemma}
To prove this lemma one again writes $B^i=G^{ij}\Sigma^j$ and looks for the connection as a sum of the self-dual part of the Levi-Civita connection compatible with the metric (\ref{Urbantke}) defined by $B^i$, and an extra term. This extra term can be computed and involves the covariant derivatives of $G^{ij}$ (with respect to the self-dual part of the Levi-Civita connection). Details can be found in e.g. \cite{Freidel:2008ku}, \cite{Krasnov:2009ik}. 

We shall call the connection in the above lemma {\it compatible} with the triple $B^i$. Note that when $B^i=F^i$ the equation (\ref{compat}) is the Bianchi identity and so is automatically satisfied. Still, (\ref{compat}) can be solved to find components of the connection algebraically expressed in terms of the derivatives of the curvature. 

The second lemma describes the solution to (\ref{compat}) in the special case that the triple $B^i$ satisfies the equation (\ref{feqs-ff}). Such triples can be called {\it perfect}.
\begin{lemma} Let $B^i$ be a perfect triple, i.e. a triple satisfying $B^i\wedge B^j\sim \delta^{ij}$. Then the connection obtained by Lemma 1 is the self-dual part of the metric (\ref{Urbantke}) compatible Levi-Civita connection.
\end{lemma}
This lemma follows from details of the proof of the previous lemma, as all terms involving the covariant derivative of $G^{ij}$ vanish in this case, and $A$ reduces to the self-dual part of the Levi-Civita connection.

Combining these two lemmas we see that when the connection satisfies (\ref{feqs-ff}) it is the self-dual part of the Levi-Civita connection compatible with the metric (\ref{Urbantke}) with $B^i=F^i$. However, by construction of the metric (\ref{Urbantke}) the curvature 2-forms $F^i$ are self-dual with respect to it. We now use the well-known decomposition of the Riemann curvature, see e.g. Besse \cite{Besse:1987pua} page 51, to conclude that (i) the metric is Einstein; (ii) the self-dual part of Weyl vanishes and so it is a gravitational instanton. The first of these follows because the curvature of the self-dual part of the Levi-Civita connection is self-dual as a 2-form. The second follows because the equation (\ref{feqs-ff}) says that there is only the scalar curvature, but no self-dual part of Weyl.  More details on the above material, together with proofs of many of the facts mentioned, can be found in e.g. \cite{Fine:2013qta}.

All in all, in the description presented here, to find a gravitational instanton one has to find an ${\rm SO}(3)$ connection satisfying (\ref{feqs-ff}), which is a first order PDE on the connection. One then gets an instanton metric by the formula (\ref{Urbantke}). The sign of the scalar curvature is determined by the sign of the triple of curvature 2-forms. Thus, if the resulting triple of curvature 2-forms $F^i$ is positive (negative) definite, see previous subsection, then the scalar curvature is negative (positive). This reversing of the sign has to do with our usage of self-dual 2-forms rather than anti-self-dual. It may well be more natural to use anti-self-dual forms, but we have refrained from doing so in this paper in order to agree with conventions of some of our earlier works. 

\subsection{Linearisation around an instanton}

We now linearise (\ref{sdgr}) around an arbitrary instanton background. Let us denote by $\Sigma^i$ the triple of orthonormal self-dual 2-forms for the background instanton metric. This triple is defined so as to satisfy the following algebra
\be
\Sigma^i_{\mu}{}^\rho \Sigma^j_{\nu\rho} = \delta^{ij} g_{\mu\nu} + \epsilon^{ijk} \Sigma^k_{\mu\nu}.
\ee
Let us for definiteness consider a positive scalar curvature instanton. Let $\Lambda$ be the cosmological constant. We define a mass scale $\Lambda/3=M^2$. Then the curvature of our background connection can be written as
\be
F^i = - M^2 \Sigma^i,
\ee
where again the minus sign here is related to our usage of self-dual 2-forms. 

As in the case of the SDYM, we first consider a background with $\Psi^{ij}=0$. A more general background is also possible, we will give some comments on this below. Denoting the field perturbations $\delta A=a, M^2 \delta \Psi^{ij}=\psi^{ij}$, where we have rescaled the $\psi$ field to give it mass dimension two, we have the following linearised Lagrangian
\be\label{L2-sdgr}
{\cal L}^{(2)} = -2 \psi^{ij} \Sigma^i d_A a^j.
\ee
There is also a cubic and a quartic interaction
\be\label{L3-sdgr}
{\cal L}^{(3)} = \frac{1}{M^2} \psi^{ij} d_A a^i d_A a^j - \psi^{ij} \Sigma^i \epsilon^{jkl} a^k a^l, \\ \nonumber
{\cal L}^{(4)} = \frac{1}{M^2} \psi^{ij} d_A a^i  \epsilon^{jkl} a^k a^l.
\ee
The wedge product of forms is everywhere implied. The quintic interaction vanishes because it makes the wedge product of at least a pair of one forms $a^i \wedge a^j$ contract in their Lie algebra index. The appearance of $1/M^2$ in the first term in the cubic interaction tells us that we have a negative mass dimension coupling constant, as is appropriate for a gravity theory. 

\subsection{Spinor description}

As in the case of self-dual YM, the Lagrangian (\ref{L2-sdgr}), (\ref{L3-sdgr}) is most clearly understood by rewriting it in terms of spinors. The tracefree field $\psi^{ij}$ becomes a rank 4 symmetric spinor $\psi^{ABCD}$, thus valued in $S_+^4$. The connection perturbation has both its spacetime index and the internal index translated into spinors. The internal ${\rm SO}(3)$ index becomes a pair of symmetrised spinor indices $(AB)$, while the spacetime index becomes a pair $MM'$. The identification of internal and spacetime spinor indices is carried out by the 2-forms $\Sigma^i$. Overall, the connection perturbation becomes an object $a_{MM'AB}$, thus valued in $S_+\times S_-\times S_+^2$. 

The linearised Lagrangian becomes
\be\label{L2-gr-spin}
{\cal L}^{(2)} = \psi^{ABCD} d_A{}^{M'} a_{BM'CD}.
\ee
Here we have absorbed some numerical factors that arise in the spinor conversion (including a factor of $\sqrt{2}$) into $\psi$. Details of our spinor conversion rules are spelled out in \cite{Delfino:2012aj}, Section 6.  

\subsection{Gauge}

Let us discuss the gauge symmetries that (\ref{L2-gr-spin}). First, there is a complete symmetrisation of the $BCD$ indices of $a_{BM'CD}$ enforced by the complete symmetry of $\psi^{ABCD}$. This means that the linearised Lagrangian is independent of some of the components of the connection. We have the decomposition
\be
S_+\times S_- \times S_+^2 = S_+^3 \times S_- \oplus S_+\times S_-.
\ee
The last term here, which contains 4 components, is what is projected out in (\ref{L2-gr-spin}). It is not hard to show, see \cite{Delfino:2012aj}, that this is precisely the part of the connection perturbation that changes under diffeomorphisms. So, a very convenient way of gauge-fixing diffeomorphisms is simply to require the connection perturbation to have only the completely symmetric $S_+^3 \times S_-$ component. This is an important point, as in the usual metric formulation the action of diffeomorphisms involves the first derivative of the gauge transformation parameter - a vector field. Therefore, one cannot disentangle the components of the metric that are pure gauge. They are usually made propagating, with appropriate ghosts added to the theory to project out the effect of the gauge modes. What happens in the connection formulation is different. The action of diffeomorphisms is algebraic, and does not involve derivatives of the gauge parameter. The components of the connection that are pure (diffeomorphism) gauge can be projected out from the start. 

However, there are also the usual ${\rm SO}(3)$ gauge transformations to deal with. The Lagrangian (\ref{L2-gr-spin}) is ${\rm SO}(3)$ gauge invariant. Indeed, when $\delta a_{MM' AB} = d_{MM'} c_{AB}$, where $c_{AB}$ is the parameter of the gauge transformation, the exterior covariant derivative applied twice gives the curvature proportional to $\Sigma^i$. When translated into spinors this gives
\be
d_A{}^{M'} d_{BM'} c^{CD} \sim \epsilon_{(A}{}^{(C} c_{B)}{}^{D)}.
\ee
This vanishes when contracted with $\psi^{AB}{}_{CD}$, which shows that (\ref{L2-gr-spin}) is ${\rm SO}(3)$ gauge invariant. 

\subsection{Gauge-fixing}

As we have already mentioned, a very convenient gauge-fixing of diffeomorphisms is to allow only the $S_+^3\times S_-$ component of the connection propagate, fixing the other part of the connection to zero. This does not require ghosts, at least not at the order quadratic in the perturbations. 

The gauge-fixing of ${\rm SO}(3)$ symmetry is done as in the Yang-Mills case, by using the gauge-fixing fermion appropriate for the Landau (sharp) gauge. The only subtlety is that we want the gauge-fixing term only depend on the $S_+^3\times S_-$ part of the connection so that only this component propagates. So, we project on this part of the connection in the gauge-fixing term. The corresponding gauge-fixing fermion is
\be\label{ferm-gr}
\Psi = \bar{c}^{AB} d^{MM'} a_{M'M AB} \equiv \bar{c}^{AB} \epsilon^{DC} d_C{}^{M'} a_{M'DAB},
\ee
where $a_{M'MAB}=a_{M'(MAB)}$ is the connection perturbation projected onto $S_+^3\times S_-$. We also wrote the gauge-fixing fermion in a suggestive form. The part of the BRST variation of (\ref{ferm-gr}) contributing to the bosonic part of the action is
\be
h_c^{AB} \epsilon^{DC} d_C{}^{M'} a_{M'DAB},
\ee
where $h_c$ is the ghost auxiliary field. As in the case of gauge-fixing the self-dual YM action, we can now combine the auxiliary field for ghosts with the field $\psi^{ABCD}$ via
\be
\tilde{\psi}^{ABCD} = \psi^{ABCD} + h_c^{AB} \epsilon^{DC}.
\ee
The new field is no longer in $S_+^4$ but rather in $S_+^3\times S_+$. This makes the full gauge-fixed quadratic part of the Lagrangian to be (\ref{L2-gr-spin}) with $\psi^{ABCD}$ replaced with $\tilde{\psi}^{ABCD}$. The tilde will now be dropped with understanding that the field $\psi^{ABCD}$ takes values in $S_+^3\times S_+$ and contains the ghost auxiliary fields. 

The operator that appears in the kinetic term is now 
\be\label{def-complex}
\dir: S_+^3\times S_- \to S_+^3 \times S_+.
\ee
This is the elliptic Dirac operator and the gauge has been fixed. All this is exactly analogous to the self-dual YM story, see (\ref{dirac-ym}). 

\subsection{Amplitudes}

We will now characterise the cubic interaction (\ref{L3-sdgr}) in terms of amplitudes that it produces. In order to do so, we need to describe the physical states. However, we encounter a complication, which is that our description of self-dual gravity only makes sense with non-zero scalar curvature. Thus, the backgrounds we can consider are not flat, with radius of curvature $1/M$. One can take the limit $M\to 0$ to approach the flat case, but in this limit the coupling constant in (\ref{L3-sdgr}) blows up, so the limit is singular. 

Fortunately, there is a systematic procedure of taking the flat limit described in \cite{Delfino:2012aj}. In this procedure one essentially works with flat space and the usual Fourier transform. However, some factors of $M$ are kept in intermediate answers, and cancel out in the (most) final expressions. 

Taking the flat limit, the linearised field equations read 
\be\label{feqs-gr-lin}
\partial_A{}^{M'} a_{BCD M'} =0, \qquad \partial_{M'}{}^D \psi_{ABCD} =0.
\ee
The operators that appear here are the usual chiral Dirac operators. They convert a spinor index of one type into the other. Squaring these operators one gets the Laplacian, which implies, when going to the momentum space, that the momentum is null. The momentum vector translated into spinors is thus a product of two spinors (\ref{k2}). 

We now describe polarisation spinors. Both negative and positive polarisations will contain factors of $M$. The solution to the first equation in (\ref{feqs-gr-lin}) is given by the following spinor
\be\label{gr-}
\epsilon^-_{ABCM'}(k) = M \frac{q_A q_B q_C k_{M'}}{\la qk\ra^3} .
\ee
The factor in the denominator is needed to make this polarisation spinor homogeneity degree zero in the reference spinor $q$. The prefactor of $M$ is needed to give it the mass dimension zero, as is appropriate for the polarisation spinor of a field of mass dimension one. A general solution to the first equation in (\ref{feqs-gr-lin}) is then a linear superposition of plane waves with such polarisation spinors. 

The other polarisation provides a solution to the second equation in (\ref{feqs-gr-lin}). It is
\be\label{gr+}
\epsilon^+_{ABCD} = \frac{1}{M} k_A k_B k_C k_D.
\ee
Again, we need a dimensionful prefactor here to get the mass dimension one, as is appropriate for a mass dimension two field. 

We can now compute the cubic vertex on shell. We insert two negative polarisation spinors for $a's$ and the positive polarisation spinor for $\psi$. The computation is simplified by noticing that in the last term in the first line of (\ref{L3-sdgr}) the commutator $\epsilon^{ijk} a^j a^k$ vanishes on two spinors of the type (\ref{gr-}), so the last term does not contribute. Also in the first term we only need to compute the ASD parts of $\partial a$ because the self-dual part vanishes by the linearised field equations. This ASD part is $M q_A q_B k_{A'} k_{B'} / \la qk \ra^2$. We then get the following result
\be
A^{--+} \sim \frac{1}{M} \frac{\la q3\ra^4 [12]^2}{\la q1\ra^2 \la q2\ra^2}.
\ee
The fraction here is just the square of (\ref{A3-ym-1}), and so using the momentum conservation we get a multiple of the square of (\ref{A3-ym-2})
\be
A^{--+} \sim \frac{1}{M} \frac{[12]^6}{[13]^2[23]^2}.
\ee
This is the familiar gravitational 3-point amplitude, apart from the fact that instead of the familiar $1/M_p$ here, $M_p$ being the Planck mass, we have $1/M$. This is because our gravitons are normalised to the scale $M$, not to the scale $M_p$. In other words, the usual gravitons are normalised so that the perturbative expansion of the metric starts as $\eta + M_p^{-1} h$, where $\eta$ is the flat metric. In the story above, the similar expansion starts as $\eta + M^{-2} \partial a$. So, our amplitudes need to be appropriately rescaled by factors of $M_p$, and in the case of the 3-point amplitude this replaces $M$ with $M_p$. These subtleties are discussed in more details in \cite{Delfino:2012aj} for the case of full GR. 

\subsection{Berends-Giele current}

Similarly to the YM case, one can compute the Berends-Giele current given by the sum of all Feynman diagrams with one off shell leg. The most interesting current is when the on shell legs are those of negative polarisation gravitons. 

As for the YM theory, we write the n-point current in a convenient form
\be\label{gr-current-form}
J_{ABCM'}(1,\ldots,n) = M^{2-n} q_A q_B q_C q^E (\sum_{i=1}^n k_i)_{EM'} J(1,\ldots,n),
\ee
where $J(1,\ldots,n)$ are now scalars. Note that the convention here is to include the final off shell leg propagator. The first current is just the polarisation spinor (\ref{gr-}) itself 
\be
J(1) = \frac{1}{\la q1\ra^4}.
\ee
Note that the form (\ref{gr-current-form}) implies that the $n$-th current remains anti-self-dual in the sense that $\partial_M{}^{M'} J_{ABCM'}=0$. This is confirmed by the computation sketched below. 

We now start building more complicated higher point currents from the lower order ones, following the Berends-Giele procedure. At second order we need to insert two order one currents into the cubic vertex. As we have already discussed in the previous subsection, we only need the first term in (\ref{L3-sdgr}) because the connection commutator in the second term vanishes on states of the type (\ref{gr-current-form}). From the first term we only need the product of the anti-self-dual parts of $da$, as the self-dual parts vanish. So, inserting two negative polarisation spinors into the cubic vertex and applying the final leg propagator we read off the result
\be
J(1,2) = \frac{1}{\la q1\ra^2 \la q2\ra^2} \frac{[12]}{\la 12\ra}.
\ee

At next order there are 3 diagrams to consider: the $J(1,2)$ current combining with $J(3)$, then $J(1)$ with $J(2,3)$ and finally $J(1,3)$ with $J(2)$. The contribution to $J(1,2,3)$ from the first of these diagrams is
\be
(q^E(k_1+k_2)_{EM'}) (q^F(k_1+k_2)_{FN'}) (q^M (k_3)_M{}^{M'}) (q^N (k_3)_N{}^{N'}) \frac{J(1,2)J(3)}{\la 12\ra [12]+\la 23\ra[23]+\la{13}\ra[13]}.
\ee
This is equal to
\be
\frac{( \la q1\ra [13]+\la q2\ra[23])^2[12]}{\la q1\ra^2\la q2\ra^2\la q3\ra^2 \la 12\ra} \frac{1}{\la 12\ra [12]+\la 23\ra[23]+\la{13}\ra[13]}.
\ee
We then add contributions from all 3 channels. After extracting a common factor, there are terms of two types. In one type we have squares of $\la q1\ra$, etc. These terms can be written as (omitting the common factor of the final propagator times $1/\la q1\ra^2\la q2\ra^2\la q3\ra^2$)
\be\label{gr-cur-1}
\la q1\ra^2 \frac{[12][13]}{\la 12\ra\la 13\ra} ( \la 13\ra[13]+\la 12\ra[12]) + \la q2\ra^2 \frac{[12][23]}{\la 12\ra\la 23\ra} ( \la 12\ra[12]+\la 23\ra[23])\\ \nonumber
+\la q3\ra^2 \frac{[13][23]}{\la 13\ra\la 23\ra} ( \la 13\ra[13]+\la 23\ra[23]).
\ee
The remaining terms are
\be\label{gr-cur-2}
\frac{[12][13][23]}{\la12\ra\la13\ra\la23\ra}\left( 2\la q1\ra\la q2\ra \la 13\ra\la 23\ra + 2\la q2\ra\la q3\ra \la 12\ra\la 13\ra-2\la q1\ra\la q3\ra \la 12\ra\la 23\ra\right).
\ee
The expression in brackets here can be transformed using the Schouten identity. Indeed, we have
\be
\la q1\ra^2 \la 23\ra^2 + \la q2\ra^2 \la 13\ra^2 +\la q3\ra^2 \la 12\ra^2  \\ \nonumber
= \left( \la q2\ra \la 13\ra - \la q3\ra \la 12\ra\right)^2+ \left( \la q1\ra \la 23\ra - \la q3\ra \la 21\ra\right)^2+\left( \la q1\ra \la 32\ra - \la q2\ra \la 31\ra\right)^2 \\ \nonumber
=2 \la q1\ra^2 \la 23\ra^2 + 2 \la q2\ra^2 \la 13\ra^2 +2 \la q3\ra^2 \la 12\ra^2 \\ \nonumber- \left( 2\la q1\ra\la q2\ra \la 13\ra\la 23\ra + 2\la q2\ra\la q3\ra \la 12\ra\la 13\ra-2\la q1\ra\la q3\ra \la 12\ra\la 23\ra\right).
\ee
This shows that
\be
2\la q1\ra\la q2\ra \la 13\ra\la 23\ra + 2\la q2\ra\la q3\ra \la 12\ra\la 13\ra-2\la q1\ra\la q3\ra \la 12\ra\la 23\ra \\ \nonumber = \la q1\ra^2 \la 23\ra^2 + \la q2\ra^2 \la 13\ra^2 +\la q3\ra^2 \la 12\ra^2 .
\ee
The terms (\ref{gr-cur-2}) can now be combined with those in (\ref{gr-cur-1}). We see that the terms in (\ref{gr-cur-2}) just provide the missing in brackets in (\ref{gr-cur-1}) terms to cancel the final propagator. The final result for the current is then
\be
J(1,2,3) = \frac{1}{\la q2\ra^2 \la q3\ra^2} \frac{[12][13]}{\la 12\ra\la 13\ra}  + \frac{1}{\la q1\ra^2 \la q3\ra^2} \frac{[12][23]}{\la 12\ra\la 23\ra} +\frac{1}{\la q1\ra^2 \la q2\ra^2} \frac{[13][23]}{\la 13\ra\la 23\ra} .
\ee
Even though more complicated than in the case of YM theory, the pattern is now becoming clear. The current is given by a sum over trees on $n$ points. A general expression can be found in e.g. \cite{Krasnov:2013wsa}. It can also be seen that the current is given by an expansion of a certain (reduced) determinant, see \cite{Krasnov:2013wsa}.

Precisely the same arguments as in the case of self-dual YM theory show that all tree-level amplitudes with $n>3$ vanish. This is because to get such amplitudes one would need to remove the final propagator that was part of the definition of the current. However, there is no pole to cancel and the resulting amplitudes are zero. 

\subsection{Quantum theory}

The discussion in this subsection parallels that in the case of YM almost verbatim, so we will be brief. Again, the fact that $\Psi$ enters the Lagrangian linearly immediately tells us that the theory is one-loop exact. To study the theory at one loop we again use the background field method formalism. 

As in the YM case, we first discuss the case of a pure instanton background with the background auxiliary field $\Psi=0$. We then make comments about the more general backgrounds. The action to study is (\ref{L2-gr-spin}), gauge-fixed as we explained above. As in the YM case we first rewrite the (bosonic part of the) Lagrangian as 
\be\label{dirac-form-gr}
{\cal L}^{(2)} = \frac{1}{2} \left( \psi \quad a  \right) \left( \begin{array}{cc} 0 & \dir \\ \dir^* & 0 \end{array} \right) \left( \begin{array}{c} \psi \\ a\end{array}\right)\equiv \frac{1}{2} \left( \psi \quad a  \right) D\left( \begin{array}{c} \psi \\ a\end{array}\right)
\ee
Here 
\be
\dir: S_+^3\times S_- \to S_+^3 \times S_+,  \qquad 
(\dir a)_{ABCD} := d_A{}^{M'} a_{BCDM'}, 
\ee
is the chiral part of the Dirac operator and 
\be
\dir^*: S_+^3\times S_+ \to S_+^3\times S_-, \qquad 
(\dir^* b)_{ABCM'} := d_{M'}{}^M b_{ABCM}, 
\ee
is its adjoint operator. The operator $D$ in (\ref{dirac-form-gr}) is the usual Dirac operator acting on 4-component spinors with values in $S_+^3$. Its determinant is computed by first squaring the operator to convert it to a Laplace type second order operator, see (\ref{dirac-square}) and then using the standard heat kernel technology. The effective action is obtained by adding contributions from the bosonic sector and the ghosts as in (\ref{s-eff}). 

\subsection{Absence of quantum divergences}

As in the case of SDYM, the logarithmic divergences are captured by the heat kernel coefficient that is of the form of the integral of curvature squared. To see if any divergence is possible on an instanton background, we need to consider the topological invariants available. In the gravity case there are two such invariants - the Euler characteristic and the signature. Both can be expressed as integrals of the appropriate curvature components squared. The Gauss Bonnet formula gives the Euler characteristic as 
\be
\chi = \frac{1}{32\pi^2} \int\left( (R_{\mu\nu\rho\sigma})^2  - 4(R_{\mu\nu})^2  + R^2 \right). 
\ee
Decomposing the Riemann curvature into its irreducible parts
\be
R_{\mu\nu\rho\sigma} = W_{\mu\nu\rho\sigma} + ( g_{\mu[\rho} Z_{\sigma]\nu} - g_{\nu[\rho} Z_{\sigma]\mu}) + \frac{R}{6} g_{\mu[\rho} g_{\sigma]\nu},
\ee
where
\be
Z_{\mu\nu}:= R_{\mu\nu}-\frac{R}{4} g_{\mu\nu}
\ee
is the tracefree part of Ricci, we get
\be
(R_{\mu\nu\rho\sigma})^2 =(W_{\mu\nu\rho\sigma})^2+ 2(Z_{\mu\nu})^2 + \frac{R^2}{6}.
\ee
Thus, we can rewrite
\be
\chi = \frac{1}{32\pi^2} \int\left( |W^+|^2+|W^-|^2 - 2 |Z|^2 + \frac{R^2}{6} \right),
\ee
where we wrote the Weyl squared as the sum of squares of its self- and anti-self-dual parts. The other topological invariant we need is the signature
\be
\tau = \frac{1}{48 \pi^2} \int \left( |W^+|^2 - |W^-|^2\right). 
\ee

Any invariant quadratic in the curvature can be decomposed into irreducible parts of the curvature squared. These parts are $W^\pm, Z, R$. On an instanton we have $Z=0, W^+=0$, so there are only two possible invariants constructed from the curvature squared. But there are also two topological numbers, and thus both of these invariants are proportional to a quantity that is topological. We have
\be
\tau = - \frac{1}{48 \pi^2} \int \left( |W^-|^2\right), \qquad 
2\chi + 3\tau = \frac{1}{16\pi^2} \int R^2.
\ee
Thus, on an instanton background any logarithmically divergent quantity is some linear combination of $\chi$ and $\tau$. These are integrals of total derivatives, and cannot contribute to the S-matrix. The theory is quantum finite, at least on an instanton backgrounds.

All this is like in self-dual YM. We should now discuss more complicated backgrounds with $\Psi\not=0$. A background of this sort modifies the operator $D$ to be considered in (\ref{dirac-form-gr}) by introducing an off-diagonal term $\Psi^{ij} d_A a^i d_A a^j$. By integration by parts this term can be written as a term of the type $d_A \Psi^{ij} a^i d_A a^j$, which contains only the first derivative of $a^i$, as well as the curvature term of the schematic type $\Psi a^2$. This modifies the operator $D$ by adding also a first order differential operator to the lower-diagonal position. This $D$ has then to be squared, and converted into an operator of Laplace type in order to use the standard heat kernel methods. Details of this has not yet been worked out. 

However, as in the YM case one can also refer to a general argument. Thus, we can argue that the one-loop effective action cannot depend on $\Psi$. Indeed, if there was such a dependence, it would imply that there are one loop amplitudes with external $\psi$ legs, but we know this is not the case. Thus, the computation of the one loop effective action can be reduced to that on $\Psi=0$ background. Then no computation is necessary, as we know before any computation that all possible divergences are topological. Self-dual gravity is thus quantum finite. 

\subsection{One loop amplitudes}

As in the YM, the all negative helicity one loop amplitudes can be conjectured based on the soft and collinear limit arguments \cite{Bern:1998xc}. One expects these amplitudes to be purely rational, because all cuts are vanishing. They can be explicitly computed (at low $n$) by using supersymmetry to replace the graviton propagating in the loop with a massless scalar, as explained in \cite{Bern:1998xc}. Nobody has computed these amplitudes from the self-dual gravity Feynman rules, in particular because no covariant formulation of SDGR was previously available. It should be possible to do this using the self-dual gravity Feynman rules described above, but this remains to be done. 

We now quote the result for the 4-point amplitude from \cite{Bern:1998xc}, see formula (17) of this reference. We omit the numerical factors. The amplitude is
\be
{\cal M}_{1-loop}^{----} \sim \frac{1}{M_p^4} \left( \frac{s_{12} s_{23}}{\la 12\ra \la 23\ra \la 34\ra \la 41\ra} \right)^2 (s_{12}^2+s_{23}^2+s_{13}^2),
\ee
where $s_{ij}=(k_i+k_j)^2$ are the usual Mandelstam variables. The general $n$ expression can be found in \cite{Bern:1998xc}. As is noted in this reference, the general $n$ all minus amplitudes in both self-dual YM and gravity exhibit the same structure. They are both built from the corresponding off shell currents in an essentially the same way, see formulas (16), (23) of this reference. 

As in the YM case, it can be conjectured that non-vanishing of these amplitudes is related to a quantum anomaly in conservation of the currents responsible for the integrability of self-dual gravity. It remains to be seen if this interpretation is correct. 

\subsection{Relation to full GR}

We now describe the relation of the above self-dual gravity theory to the full GR. As in the case of YM, the full theory is obtained by adding to the Lagrangian extra terms. Unlike the YM case, in the case of gravity one needs an infinite number of such terms.  

The action for full GR in this language reads
\be\label{full-gr}
S^{GR}[A,\Psi] = - \frac{M_p^2}{2M^2} \int \left( (\id + \Psi)^{-1}\right)^{ij} F^i \wedge F^j = 
\frac{M_p^2}{2M^2} \int \left( - \id + \Psi - \Psi^2 + \ldots \right)^{ij} F^i \wedge F^j.
\ee
The first term here is topological. The second term is a multiple of our self-dual gravity action (\ref{sdgr}). The third and the following terms are new. The new terms modify the propagator by making also $\la aa\ra$ non-zero, which is like in the case of YM theory. The new terms also add an infinite number of new interaction vertices. We note that as in the metric GR, the vertices that follow from (\ref{full-gr}) contain at most two derivatives. 

The action (\ref{full-gr}) is non-polynomial in one of the fields, namely in $\Psi$. This is no better in the usual metric formulation of GR, where the action is non-polynomial in the metric due to the inverse metric explicitly appearing in its construction. In this sense (\ref{full-gr}) is no worse than the usual Einstein-Hilbert action. 

The above action is normalised to coincide with
\be
S^{GR}[g]= -\frac{\Lambda}{8\pi G} \int \sqrt{g}
\ee
on Einstein metrics. In (\ref{full-gr}) $M^2=\Lambda/3$ and $M_p^2 = 1/8\pi G$. 

A quick explanation of why (\ref{full-gr}) is the correct gravity action is as follows. It is very convenient to introduce the notation 
\be\label{Sigma-def}
\Sigma^i := \left( (\id + \Psi)^{-1}\right)^{ij} F^j.
\ee
The field equations are then as follows. When varying with respect to $\Psi^{ij}$ we get 
\be
\Sigma^i \wedge \Sigma^j \sim \delta^{ij}.
\ee
Varying with respect to the connection we get
\be
d_A \Sigma^i =0.
\ee
These are the familiar equations from our discussion of the self-dual gravity. Both equations together imply that $A^i$ is the self-dual part of the Levi-Civita connection compatible with the metric defined by $\Sigma^i$ via (\ref{Urbantke}), see Lemmas 1,2 in subsection 3.3. Then (\ref{Sigma-def}) implies that the curvature of the self-dual part of the Levi-Civita connection is self-dual as the 2-form. This is equivalent to the Einstein condition, as follows from the decomposition of the Riemann curvature, see \cite{Besse:1987pua} page 51. The equation (\ref{Sigma-def}) also identifies $\Psi^{ij}$ as a multiple of the self-dual part of the Weyl curvature. 

The auxiliary fields $\Psi^{ij}$ can be integrated out from (\ref{full-gr}) to obtain a "pure connection" description of GR. The corresponding Lagrangian is analogous to the usual $F^2$ Lagrangian for YM. The pure connection formulation of GR was first described in \cite{Krasnov:2011pp} and is reviewed from a more mathematical perspective in \cite{Fine:2013qta}. We note that results in particular in \cite{Herfray:2015fpa} suggest that it may be more useful to work in the formulation (\ref{full-gr}), and not in the formulation with $\Psi$ integrated out, also for the full GR. For the self-dual gravity one does not have an option, as only the first order description with field $\Psi$ is possible.

Because of the relation (\ref{full-gr}) between the self-dual and full gravity theories we can immediately conclude that some of the amplitudes of the full GR are correctly captured by the simpler self-dual gravity theory. It can then be anticipated, in analogy to \cite{Bardeen:1995gk}, that integrability of the self-dual gravity may be of help to understand the full gravity theory. Whether this point of view can be useful remains to be seen. 

\subsection{Another covariant formulation}

There exists another covariant formulation of self-dual gravity \cite{Herfray:2015rja}. It contains more fields, but allows to describe the flat case instantons as well, unlike the above connection formulation. However, this other formulation has the drawback of being much more non-linear than (\ref{sdgr}). In particular, it is no longer linear in one of the fields, and so the already used argument for one loop exactness of this theory is no longer valid. So, it is not clear if this formulation is a good starting point for the construction of the quantum theory. We refrain from giving details here, referring an interested reader to  \cite{Herfray:2015rja}.

\subsection{Twistor space description}

The fact that in our formalism the instanton condition is first order in derivatives suggests that there is a link to the twistor description, where the instanton condition takes the form of Cauchy-Riemann holomorphicity equation, which is also first order. This turns out to be true, there is a direct link between the twistor space description of instantons and (\ref{sdgr}). Details of this relation have been worked out in \cite{Herfray:2016qvg}, where we refer the reader for more details. 

To describe it, we note that the connection $A^i$, which we will now view as an ${\rm SU}(2)$ connection $A^{AB}$, defines a set of 1-forms in the $\C^2$ bundle over the space. Let $\pi_A$ be the (holomorphic) coordinates on $\C^2$. Then we have
\be
D\pi_A := d\pi_A + A_A{}^B \pi_B.
\ee
There are also the complex conjugate one-forms. In turn, the connection can be defined geometrically as the 4-dimensional distribution that is in the kernel of $D\pi_A$ and its complex conjugate. 

In twistor theory one works with the projective twistor space. To pass to this we consider the Euler vector fields $E=\pi^A \partial/\partial \pi^A, \hat{E} = \hat{\pi}^A \partial/\partial \hat{\pi}^A$. For our notation on the hat operator, as well as our spinor notations see the Appendix. The forms in the total space of the $\C^2$ bundle that vanish when $E,\hat{E}$ are inserted descend to the projective twistor space. Such a form $\alpha$ is of a definite degree of homogeneity $O(n,m)$ if ${\cal L}_E \alpha = n \alpha, {\cal L}_{\hat{E}} \alpha = m\alpha$. 

From the two components of the 1-form $D\pi_A$ that can be chosen to be $\hat{\pi}^A D\pi_A, \pi^A D\pi_A$, only the second one descends to the projective twistor space. Thus, we define
\be
\tau := \pi^A D \pi_A.
\ee
It descends to a form of degree $O(2,0)\equiv O(2)$, as is easy to check. 

Another simple computation gives
\be\label{d-tau}
d\tau = D\pi^A D\pi_A + \pi^A F_A{}^B \pi_B,
\ee
where $F_A{}^B= dA_A{}^B + A_A{}^C A_{}^B$ is the curvature of our connection $A^{AB}$. The first term in $d\tau$ does not descend to the projective twistor space, as is easy to see using the identity
\be
\epsilon_A{}^B = \frac{\pi_A \hat{\pi}^B - \hat{\pi}_A \pi^B}{ \la \hat{\pi} \pi\ra}.
\ee
Inserting this decomposition of the identity into the first term in (\ref{d-tau}) gives
\be
D\pi^A D\pi_A = \frac{ 2 \hat{\pi}^A D\pi_A \wedge \tau}{ \la \hat{\pi} \pi\ra}.
\ee
This contains $\hat{\pi}^A D\pi_A$ that gets projected to zero in $\PT$. Thus, $d\tau$ in $\PT$ is just the curvature. Note that $d\tau$ projected to $\PT$ is a form of degree $O(2)$. 

One can then forget about the connections and consider $O(2)$-valued 1-forms in the projective twistor space $\PT$. To define the notion of degree of homogeneity one needs to assume that $\PT=M\times S^2$. We can now write the following action, see \cite{Mason:2007ct}
\be\label{twistor-action}
S[\psi, \tau] = \int_{\PT} \psi\wedge \tau\wedge d\tau\wedge d\tau.
\ee
Here in order for the integral over the fiber to make sense $\psi$ must be a form of homogeneity degree $O(-6)$. Clearly $\psi$ must also be a 1-form in the total space. We do not have any pre-defined notion of the complex structure in our projective twistor space, and so we cannot require any holomorphicity properties from $\psi$. 

Varying (\ref{twistor-action}) with respect to $\psi$ one gets 
\be
\tau\wedge d\tau \wedge d\tau =0.
\ee
In view of (\ref{d-tau}), this is clearly equivalent to
\be
F^{AB} F^{CD} \pi_A \pi_B \pi_C \pi_D =0 \quad \Leftrightarrow \quad F^{(AB} \wedge F^{CD)}=0.
\ee
This is precisely our condition (\ref{feqs-ff}) written in the spinor language. 

It is also clear that the above action corresponds to our spacetime action (\ref{sdgr}), with $\Psi$ given by the Penrose transform of $\psi$
\be
\Psi_{ABCD} = \int_{\CP^1} \pi_A \pi_B \pi_C \pi_D \, \psi\wedge \tau.
\ee
More details on this twistor description can be found in \cite{Yannick}.

\section*{Acknowledgments}

The author was supported by ERC Starting Grant 277570-DIGT\@ and is grateful to Evgeny Skvortsov for asking a question about self-dual gravity that led to this paper. The author is grateful to Yannick Herfray for many fruitful discussions about self-dual gravity, and for reading a draft of this paper. It is also important to acknowledge that the idea to expand in power series in (\ref{full-gr}) belongs to Yannick. The author is grateful to W. Siegel for attracting his attention to \cite{Siegel:1992wd}.

\section*{Spinors}

The aim of this Appendix is to establish our spinor notations. For concreteness, we only describe the Riemannian signature case. In this case the 4 coordinates of $\R^4$ can be collected into a matrix 
\be\label{x}
{\bf x} = \left( \begin{array}{cc} \alpha & \beta \\ -\beta^* & \alpha^* \end{array}\right) = \left( \begin{array}{cc} x_1+\im x_2 & x_3+\im x_4 \\ -x_3+\im x_4 & x_1-\im x_2 \end{array}\right).
\ee
It is clear that 
\be\label{norm}
{\rm det}({\bf x}) = |\alpha|^2+|\beta|^2 = x_1^2+x_2^2+x_3^2+x_4^2
\ee
is the usual norm of a vector in $\R^4$. We can alternatively write the flat $\R^4$ metric as
\be\label{metric-flat}
ds^2 = \frac{1}{2}{\rm Tr}( d{\bf x} d{\bf x}^\dagger).
\ee
The group of rotations ${\rm SO}(4)$, or rather its double cover ${\rm SU}(2)\times {\rm SU}(2)$ acts on $\R^4$ as
\be\label{x-transf}
{\bf x} \to g_L {\bf x} g_R^\dagger, \qquad g_{L,R}\in {\rm SU}(2).
\ee
There are then two types of spinors. We have the so-called unprimed $\lambda_A$ (primed $\lambda_{A'}$) spinors that transform as the fundamental representation of ${\rm SU}(2)_L$ (${\rm SU}(2)_L$)
\be
\lambda_A \to (g_L)_A{}^B \lambda_B, \quad \lambda_{A'} \to (g_R)_{A'}{}^{B'} \lambda_{B'}.
\ee
We shall refer to unprimed spinors as taking values in $S_+$, and to primed spinors as taking values in $S_-$. 

The bilinear form in the space of both types of spinors is given by
\be\label{pairing}
\la \lambda \eta \ra := \lambda^T \epsilon \eta,
\ee
where
\be
\epsilon = \left( \begin{array}{cc} 0 & 1 \\ -1 & 0 \end{array}\right).
\ee
The row $\lambda^A:=\lambda^T \epsilon$ can be referred to as the spinor with its index raised, so that the spinor contraction takes the form $\la \lambda \eta \ra = \lambda^A \eta_A$. Because of the property $\epsilon g^\dagger = g^T \epsilon$ that is valid for any $g\in {\rm SU}(2)$ we see that the spinors with raised index transform as
\be
\lambda^T \epsilon \to \lambda^T g^T \epsilon = \lambda^T \epsilon g^\dagger.
\ee
Thus, we see that the matrix ${\bf x}$ transforms (\ref{x-transf}) as an object ${\bf x}_{A}{}^{A'}$ with a spinor index of one type down and the other type up. 

The pairing (\ref{pairing}) of a spinor with itself vanishes. There is however another, hermitian pairing given by
\be
\la \hat{\lambda} \eta \ra := (\lambda^*)^T \eta.
\ee
This pairing applied to a spinor with itself is non-vanishing. It also introduces an anti-linear map
\be
\hat{\lambda}=\epsilon \lambda^*.
\ee
Note that the hatted spinor transforms in just the same way as the unhatted one
\be
\hat{\lambda}=\epsilon \lambda^* \to \epsilon g^* \lambda^* = g \epsilon \lambda^* = g\hat{\lambda}.
\ee
Thus, the operation $\hat{\cdot}$ maps the spaces $S_\pm$ into themselves
\be
\hat{} : S_\pm = S_\pm.
\ee
It is also easy to see that the operation hat squares to minus the identity
\be
\hat{\cdot}^2 = - \id.
\ee

Using this anti-linear map we can give another characterisation of matrices (\ref{x}). These can be characterised as those with the property $\epsilon {\bf x}^* \epsilon^T = {\bf x}$. However, this property is just the statement that the operation $\hat{\cdot}$ applied to both spinor indices of ${\bf x}$ leaves this object unchanged. So, matrices of the type (\ref{x}) are {\it real} objects in  $S_+\times S_-$ in the sense of the hat operation. Note that there are no real objects in $S_\pm$. 

It is also convenient to use objects with both indices down ${\bf x}_{AA'}$ (or up). Doing so, one translates every vector (or one-form) into an object with two spinor indices of different types. Let us rewrite the metric (\ref{metric-flat}) in the spinor notations. First, using $d{\bf x}^\dagger = \epsilon^T d{\bf x}^T \epsilon= -\epsilon d{\bf x}^T \epsilon$ we see that the matrix $d{\bf x}^\dagger$ is just {\it minus} the matrix $d{\bf x}_A{}^{A'}$ with index $A$ raised and $A'$ lowered. And so
\be
ds^2 = -\frac{1}{2} d{\bf x}_A{}^{A'} d{\bf x}^A{}_{A'} = \frac{1}{2} d{\bf x}_{AA'} d{\bf x}^{AA'}.
\ee

\end{document}